\documentclass[twocolumn]{aastex631}

\usepackage{diagbox}

\newcommand{\xzero}{x_0}
\newcommand{\yzero}{y_0}
\newcommand{\Dzero}{D_0}
\newcommand{\vxzero}{v_{x,0}}
\newcommand{\vyzero}{v_{y,0}}
\newcommand{\vzero}{V_0}

\begin{document}

\title{Death by Impact: \\Evidence for Merger-Driven Quenching in a Collisional Ring Galaxy at Cosmic Noon}

\author[0009-0009-6563-282X]{Amir H. Khoram}
\affiliation{Dipartimento di Fisica e Astronomia, Università di Bologna, Via Gobetti 93/2, I-40129, Bologna, Italy}
\affiliation{INAF, Astrophysics and Space Science Observatory Bologna, Via P. Gobetti 93/3, I-40129 Bologna, Italy}

\author[0000-0002-5615-6018]{Sirio Belli}
\affiliation{Dipartimento di Fisica e Astronomia, Università di Bologna, Via Gobetti 93/2, I-40129, Bologna, Italy}

\author[0000-0003-3121-6616]{Carlo Nipoti}
\affiliation{Dipartimento di Fisica e Astronomia, Università di Bologna, Via Gobetti 93/2, I-40129, Bologna, Italy}

\author[0000-0002-6389-6268]{Raffaele Pascale}
\affiliation{INAF, Astrophysics and Space Science Observatory Bologna, Via P. Gobetti 93/3, I-40129 Bologna, Italy}

\author[0000-0001-7769-8660]{Andrew B. Newman}
\affiliation{The Observatories of the Carnegie Institution for Science, 813 Santa Barbara St., Pasadena, CA 91101, USA}

\author[0000-0003-3816-7028]{Federico Marinacci}
\affiliation{Dipartimento di Fisica e Astronomia, Università di Bologna, Via Gobetti 93/2, I-40129, Bologna, Italy}
\affiliation{INAF, Astrophysics and Space Science Observatory Bologna, Via P. Gobetti 93/3, I-40129 Bologna, Italy}

\author[0000-0001-7782-7071]{Richard S. Ellis}
\affiliation{Department of Physics and Astronomy, University College London, Gower Place, London WC1E 6BT, UK}

\author{Letizia Bugiani}
\affiliation{Dipartimento di Fisica e Astronomia, Università di Bologna, Via Gobetti 93/2, I-40129, Bologna, Italy}
\affiliation{INAF, Astrophysics and Space Science Observatory Bologna, Via P. Gobetti 93/3, I-40129 Bologna, Italy}

\author{Matteo Sapori}
\affiliation{Dipartimento di Fisica e Astronomia, Università di Bologna, Via Gobetti 93/2, I-40129, Bologna, Italy}
\affiliation{INAF, Astrophysics and Space Science Observatory Bologna, Via P. Gobetti 93/3, I-40129 Bologna, Italy}

\author[0000-0002-3818-1746]{Eric Giunchi}
\affiliation{Dipartimento di Fisica e Astronomia, Università di Bologna, Via Gobetti 93/2, I-40129, Bologna, Italy}

% \collaboration{20}{(AAS Journals Data Editors)}

\begin{abstract}
The role of interactions and mergers in the rapid quenching of massive galaxies in the early Universe remains uncertain, largely due to the difficulty of directly linking mergers to quenching. Collisional ring galaxies provide a unique opportunity, as their morphology allows precise dating of the interaction, which can then be compared to quenching timescales inferred from star formation histories. We study a gravitationally bound system at $z=1.61$ in the UDS field, composed of a Host galaxy ($M_\star = 10^{11.4} \, M_\odot$) with a collisional ring and an X-ray AGN, and the Bullet galaxy ($M_\star = 10^{11.2} \, M_\odot$), located at a projected distance of $\sim 8$ kpc. Combining JWST and HST imaging with Keck/MOSFIRE spectroscopy, we find compelling evidence for an ongoing starburst in the Host concurrent with rapid quenching in the Bullet. The ring, $\sim 20$ kpc in diameter, is expanding at $\mathrm{127^{+72}_{-29}}$ km s$^{-1}$, implying the galaxies first collided 47--96 Myr ago. This timeline is consistent with the Host’s current starburst and the Bullet’s sudden quenching, strongly suggesting both phenomena were triggered by the interaction. Crucially, the Bullet shows no evidence of a preceding starburst, ruling out rapid gas consumption as the primary quenching channel. Instead, we suggest that merger-driven processes—such as enhanced turbulence and disk instabilities—may have suppressed star formation. An additional possibility, which we term the \textit{Dragon Effect}, is that AGN-driven outflows from the Host disrupted the Bullet’s low-density molecular gas, thereby preventing efficient star formation and accelerating quenching.

\end{abstract}


\section{Introduction} \label{sec:intro}

Massive quiescent galaxies observed in the early Universe appear to have undergone rapid cessation of star formation, yet the physical mechanisms responsible for this transition remain a topic of active debate \citep[][and references therein]{2018Man}. A central question is whether all quiescent galaxies experience a starburst phase prior to quenching, and what processes govern such rapid evolutionary changes. The epoch known as Cosmic Noon -- characterized by peak levels of star formation, quiescence, and galaxy mergers -- provides a crucial window for investigating these dynamics \citep[e.g.,][]{2014Madau,2014Conselice,2020AForsterSchreibe}. Although active galactic nucleus (AGN) feedback is frequently proposed as the primary driver of quenching \citep[e.g.,][]{2005DiMatteo,2008MNRASomerville,2015Erb,2017Beckmann,2024Belli,2024Park,2025Bugiani}, other studies highlight the significant roles of gas depletion and merger-induced perturbations \citep[e.g.,][]{2024Ellison,2025Suess,2025Gordon,2025Hewitt}. 

Major, gas-rich mergers have been proposed as an important driver of galaxy transformation, potentially operating through multiple pathways. They trigger both quasar activity and starbursts, and subsequently quench star formation through feedback-driven gas removal \citep{2008Hopkins}. Alternatively, major mergers may lead to a highly turbulent interstellar medium (ISM) in post-starburst, early-type galaxies, reducing star formation efficiencies by up to two orders of magnitude \citep[e.g.,][]{2020MNRASKretschmer}. Additionally, several studies have suggested that major mergers can suppress star formation efficiency by disrupting the gravitational collapse of gas clouds \citep[e.g.,][]{2018Ellison,2018vandeVoort,2025Suess}. This suppression is often linked to violent disc instability—a dynamical phenomenon that can be triggered in gas-rich major mergers—resulting in elevated ISM turbulence and hindered star formation \citep[e.g.,][]{2014Dekel&Burkert,2015Zolotov}, even in the absence of a starburst. Observational studies have discovered several quenched galaxies involved in major mergers, including in the early universe \citep[e.g.,][]{2018Schreiber,2022Runco}. However, it is often difficult to establish whether the quiescent state of the galaxy is indeed physically linked to the ongoing merger.

The difficulty of establishing a direct causal link between mergers and quenching can be overcome by deriving a precise timeline of both the merger and the quenching process, and then testing whether they are closely associated. This is possible only for special cases. In particular, Collisional Ring Galaxies \citep[hereafter CRG, ][]{1976Lynds,1978Toomre,1993Hernquist}, which form as a result of mergers, offer precise observational insight into the timeline of the galaxy interaction through detailed analyses of their radially propagating density waves. 
One of the most prominent examples of CRGs in the local Universe is the Cartwheel Galaxy \citep{1977Fosbury,1992Marcum}, widely regarded as the prototypical case of this phenomenon \citep[see also,][]{1992Marcum,1995Higdon,2003Gao,2004Wolter,2005Mayya,2020Murugeshan,2023Hosseinzadeh}. Recent results by \citet{2024Ditrani} show that the star formation rate increases with distance from the center, peaking in the outer ring, while the gas-phase metallicity decreases with radius \citep{2022Zaragoza-Cardiel}. Both trends in SFR and metallicity are consistent with the typical radial behavior observed in disc galaxies \citep[e.g.,][]{1982Guesten,2012Kennicutt,2014Sanchez,2025Khoram}, confirming that the collisional ring is formed from the gas in the outer disk. Moreover, \citet{2023Mayya} report that the UV-bright regions within the Cartwheel's ring host multiple stellar populations, with the bulk of the far-ultraviolet emission arising from nonionizing stars aged between approximately 20 and 150 Myr. Note that the Cartwheel system is not an isolated case but instead exemplifies a broader class of CRGs in the local Universe.

Other studies on CRGs, both in observations and simulations, exhibit enhanced star formation in the ring, typically characterized by a quiescent/low-SFR central region surrounded by a vigorously star-forming ring \citep[e.g.,][]{2005Mayya, 2007Bizyaev, 2020Murugeshan, 2024Ditrani}. This trend was specifically observed by \citet{2020Murugeshan} through the analysis of $\mathrm{H\textsc{i}}$ distribution, which in most cases revealed a significant concentration of neutral hydrogen in the ring and a corresponding deficiency in the central regions of CRGs. The observed ring expansion velocities in CRGs vary widely, ranging from a few tens to nearly 250 km/s \citep[e.g.,][]{2011Conn,2015Parker,2025Pasha}. In previous studies, reconstructing the radial expansion velocity of the ring has enabled estimates of its age and, consequently, the timing of the initial collision—typically within a few hundred million years. Additionally, tidal tails—frequently associated with major mergers in both the nearby and distant Universe \citep[e.g.,][]{2016Wen,2021Elmegreen}—are present in a subset of these galaxies, offering morphological evidence of recent gravitational interactions \citep[e.g.,][]{2011Conn,2018Renaud,2023AMartinez-Delgado,2025Pasha}. 

Both models and simulations consistently indicate that star-forming rings linked to tidal structures in CRG systems \citep[e.g.,][]{2023AMartinez-Delgado} have relatively short lifetimes, typically lasting only 0.2–0.5 Gyr \citep[e.g.,][]{2006Wong,2018Renaud}, and become only faintly detectable, in the local Universe, up to approximately 0.7 Gyr after the initial collision \citep{2015Wu,2018Elagali}. Moreover, \cite{2012FiacconiF} showed that the ring’s expansion velocity is mainly governed by the mass of the intruding galaxy. Additionally, changes in the SFH can begin as early as 50 Myr prior to the collision, followed by an increase in the Host galaxy’s SFR after the impact \citep[e.g.,][]{2018Renaud}.

In this work, we identify a gravitationally bound interacting system at $z=1.61$ consisting of an AGN-hosting CRG and a massive quiescent companion based on visual inspection of publicly available JWST/NIRCam imaging. By combining high resolution JWST and HST imaging with spatially resolved Keck/MOSFIRE spectroscopy in the Y and H band, we investigate the physical and kinematic properties of the system in detail. This multi-wavelength dataset enables us to derive the SFHs of distinct regions and compare them, for the first time at high redshift, to the timeline of the galaxy collision. Our findings reveal compelling evidence of merger-driven transformation: a starburst episode in the AGN-hosting ring galaxy and simultaneous quenching in the companion.  This system presents a rare opportunity to examine how SFHs respond to dramatic galactic perturbations driven by major mergers and AGN activity during cosmic noon.

The paper is organized as follows. In Section~\ref{sec:system} we introduce the system, followed by a detailed analysis of the spatially resolved photometry in Section~\ref{sec:photometry}. In Section~\ref{spectral_analysis} we use the spectroscopy to investigate the kinematics and SFHs. We then connect the observed physical transformations in the system to the merger event, and further discuss the implications in Section~\ref{sec:discussion}. A summary of the main findings is presented in Section~\ref{conclusion}. Throughout this paper, we adopt the \textit{Planck} 2020 cosmology \citep{2020Planck}.

\section{The System}\label{sec:system}

\begin{figure}
    \centering
    \includegraphics[width=0.95\linewidth,height=\textheight,keepaspectratio]{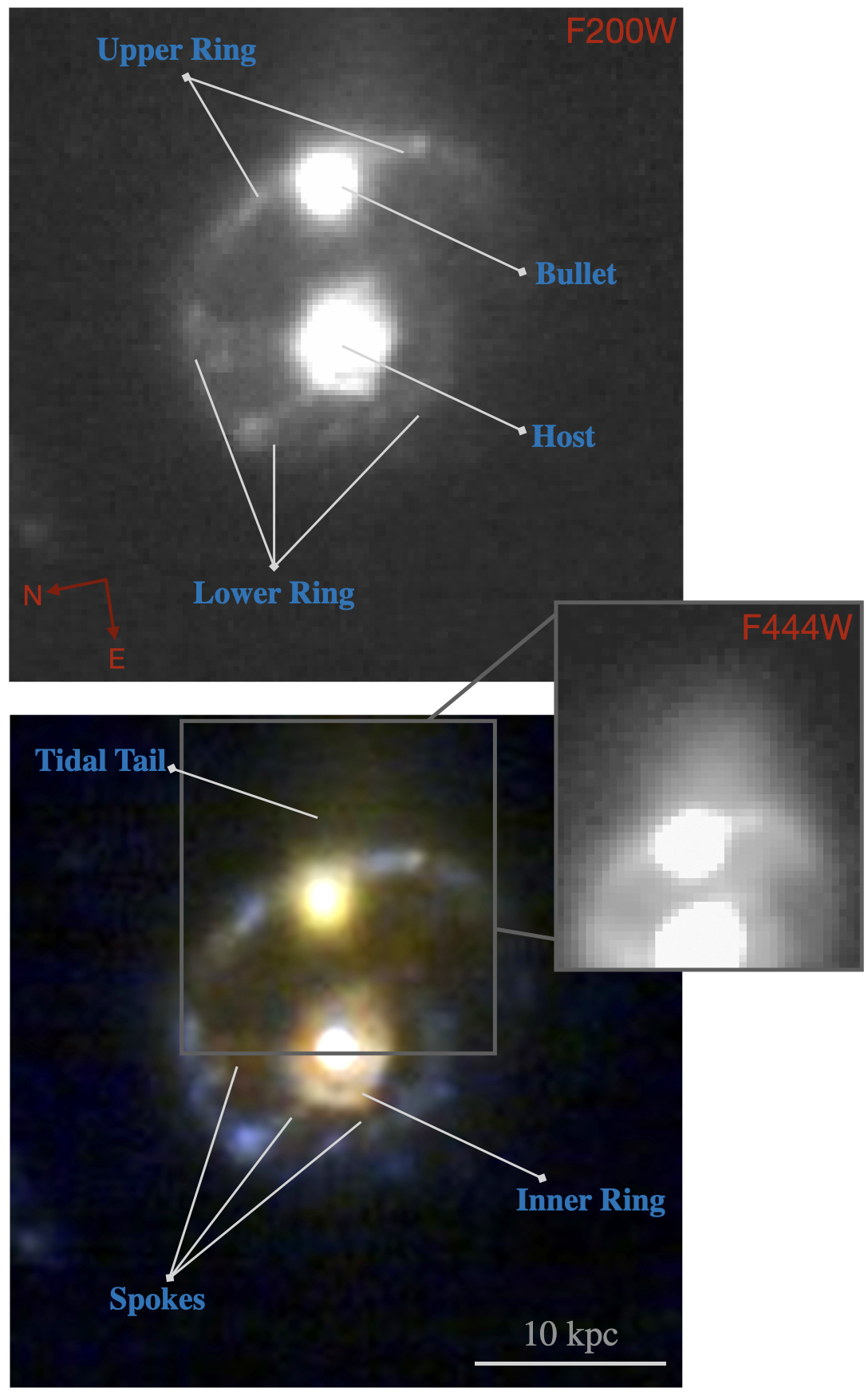}
    \caption{JWST/NIRCam images of the system. The upper panel shows the F200W-band image with labels indicating prominent components for which we have kinematic measurements. The lower panel presents a composite color image using the F090W, F115W, F150W, and F200W NIRCam bands. The tidal tail, highlighted in the inset, is more evident in the F444W filter.}
    \label{fig:system}
\end{figure}

We identified an interacting system in the UDS field at redshift 1.61 (see Fig.~\ref{fig:system}) through visual inspection of publicly available JWST broad-band imaging. The axisymmetric ring structure, along with bright spokes connecting the ring to its central galaxy (UDS 35606, hereafter the Host), are hallmark features of CRGs. The presence of an intruder galaxy (UDS 35616, referred to as the Bullet) at the same redshift, together with a prominent tidal tail, provides strong evidence supporting this classification. Based on dynamical analysis (see Appendix~\ref{Appx:fate of the system}), the two galaxies are gravitationally bound and will eventually merge. Spectroscopic confirmation for both the Host and the Bullet is presented in Sec.~\ref{spectral_analysis}. This system stands out as a rare high-redshift CRG candidate, offering a valuable opportunity for detailed study.

\begin{table}[ht]
\centering
\begin{tabular}{c|c|c}
\textbf{Region} & \textbf{$\mathrm{log\, (M_\star/M_\odot)}$} & \textbf{RA/Dec} \\
\hline
Bullet (UDS 35616) & $11.1\pm0.2$ & 34.4307660/-5.1577951 \\
Tidal tail& $10.7\pm0.2$  \\ \cline{1-2}
total&  $11.2$ \\
\hline
Host (UDS 35606)& $10.7\pm0.2$ & 34.4310037/-5.1577292 \\
Inner Ring&$10.9\pm0.3$  \\
Lower Ring& $10.5\pm0.4$   \\
Upper Ring& $10.7\pm0.2$   \\ \cline{1-2}
total& 11.4  \\
\hline
\end{tabular}
\caption{Stellar mass estimates of different regions within the system, derived from JWST/NIRCam and HST photometric data. The total stellar masses of the Host and the Bullet are obtained by summing the masses of regions spatially associated with each galaxy. Similarly, the stellar masses of the Lower and Upper Ring structures are calculated by summing the masses of their respective associated regions. The coordinates (in degrees) of the Host and the Bullet are provided in the last column.}
\label{tab:Mass_RA_Dec}
\end{table}

Figure~\ref{fig:system} illustrates various morphological features of the system, using JWST/NIRCam imaging at 2.0 and 4.44$\mathrm{\mu m}$, as well as an RGB composite. The Bullet is located at a projected distance of $\sim10$ kpc from the Host and is visually aligned with the Upper Ring in the background (see Sec.~\ref{subsec:kinematics}). Distinct spokes are observed in the F200W-band image, connecting the ring structure to the central Host galaxy. The brightest parts of the ring are more prominent in the bluer bands, while the extended tidal tail is more easily distinguishable in the redder wavelengths. Notably, the Host galaxy is X-ray detected \citep[\textit{Chandra} observations;][]{2018Kocevski_UDSXRay} and classified as a powerful AGN, with an X-ray luminosity of $L_X \sim 10^{44}$ erg\,s$^{-1}$.

For this system, eight JWST/NIRCam broad and medium bands\footnote{F090W, F115W, F150W, F200W, F270W, F356W, F410M, F444W.} are available from the PRIMER survey (GO 1837; PI: J. Dunlop). Additionally, we utilize mosaics from the 3D-HST survey, covering five HST bands\footnote{F606W, F814W, F125W, F140W, F160W.} from the ACS and WFC3 instruments \citep[][]{2011Grogin,2011Koekemoer,2014Skelton}. The images in each band are resampled and smoothed with a Gaussian kernel accounting for the spatial resolution and the PSF of the reddest band of the same telescope---F160W for HST and F444W for JWST.

We also make use of Keck/MOSFIRE spectra in the Y and H bands from \citet{2017Belli}.
The observations were conducted with a $0.7''$ slit width, yielding a spectral resolution of \( R \sim 3500 \), with exposure times of 5.5 hours in Y and 1.5 hours in H, in a $0.8''$-seeing. At the redshift of our target, in the rest frame, the Y band optimally captures key absorption features around the Balmer break ($3700$–$4200\ \text{\AA}$), while the H band ($5600$–$6700\ \text{\AA}$) contains essential emission lines, including $\mathrm{H\alpha}$, $\mathrm{[N \,II]\lambda6548\AA}$, and $\mathrm{[N\, II]\lambda6584\AA}$, which are critical for SFR estimation and classification of the ionizing source. The spectra are analyzed in Section~\ref{spectral_analysis}.

\section{SED Fitting }\label{sec:photometry}

 \begin{figure*}[ht]
    \centering
    \includegraphics[width=1.\linewidth]{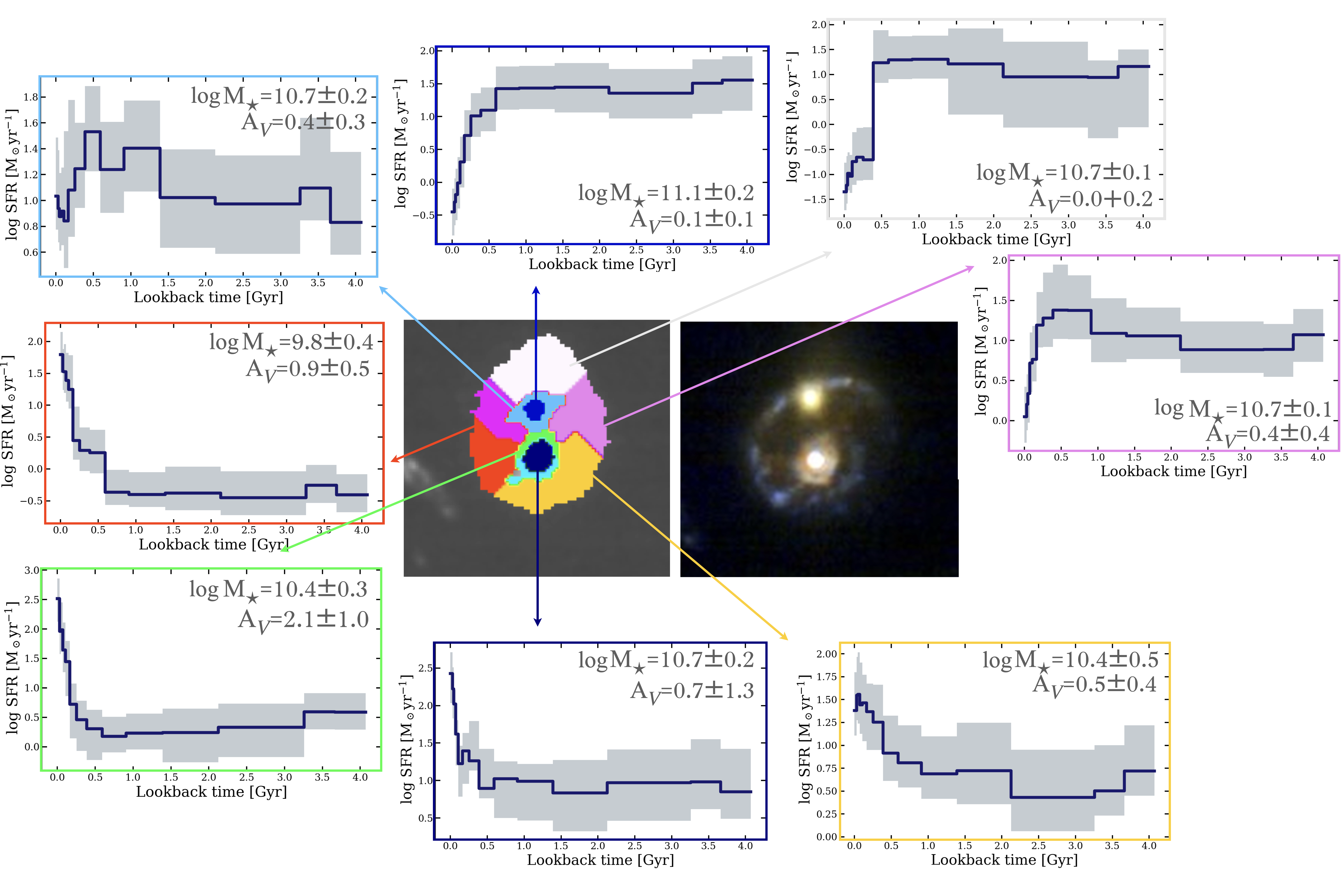}
    \caption{Star formation histories of different regions in the merging system, derived from JWST/NIRCam and HST photometric data and modeled using \texttt{\textsc{prospector}}. The inferred dust attenuation and stellar mass (in M$_\odot$) for each region are indicated within the corresponding panel. The regions are color-coded and overlaid on a NIRCam F200W cutout (left image), with the corresponding RGB image (right image) providing a visual reference for their locations within the system. Each region is connected with an arrow to its SFH panel, which is placed in a color-coded box matching the region’s color to aid visual interpretation. Note that regions are identified using segmentation map as described in Sec.~\ref{sec:photometry}. For visualization purposes, we exclude the SFHs of two regions that are heavily contaminated by neighboring sources, while, their $\mathrm{M_\star}$ are included in the total values reported in Table~\ref{tab:Mass_RA_Dec}. }
    \label{fig:region_sfh}
\end{figure*}

We identify several regions in the system, shown in the central panel of Fig.~\ref{fig:region_sfh}, using the segmentation map generated by the \texttt{photutils} Python library \citep{larry_bradley_2024_13989456}.   Given the broad color range of the system (see Fig.~\ref{fig:system}), the segmentation map is constructed from a stacked combination of all NIRCam images to ensure that all bright regions are included. A region is considered valid if it contains at least ten contiguous pixels, lying above 30$\sigma$ relative to the background noise level, thereby ensuring significant detections. The total flux for each photometric band is computed by summing the pixel values within each detected region. The associated uncertainty is derived from two main components: the pixel-wise flux errors obtained from the weight map (combined in quadrature across the region), and the contribution from background flux. To estimate the background, sources are first masked using a 2$\sigma$ detection threshold over the background to exclude faint objects. The median pixel value of the remaining background is then adopted as the nominal background level, and is multiplied by the number of pixels within each region to estimate the total background contamination. Finally, the background contribution and the weight-map-based flux error are added in quadrature. To account for potential systematic uncertainties, a 5-percent systematic error is added to the final uncertainty.

\begin{figure*}
    \centering
    \includegraphics[width=0.88\linewidth]{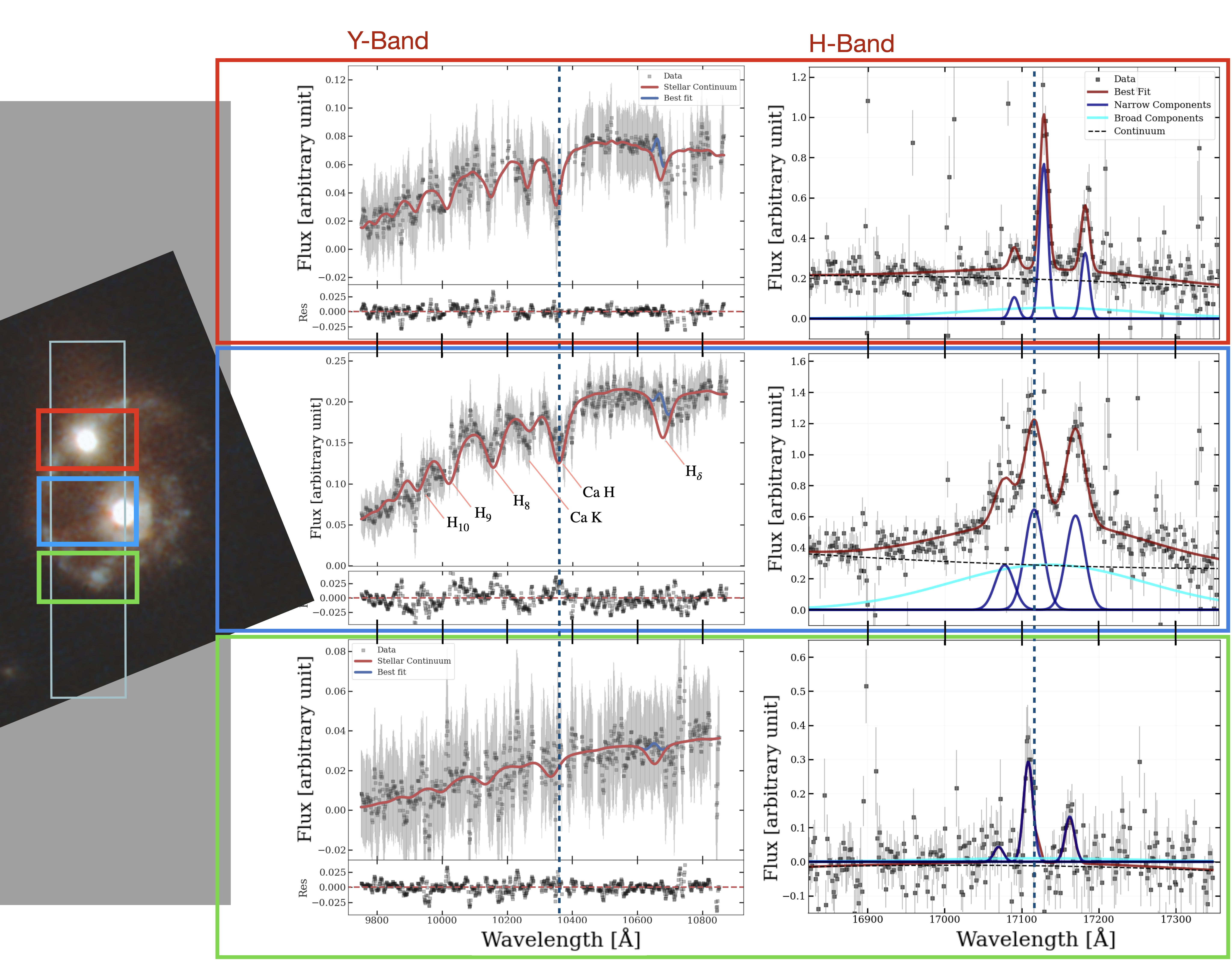}
    \caption{Keck/MOSFIRE Y  and H band spectra extracted from three spatial regions along the slit: the Bullet and Upper Ring (top), the Host galaxy and Inner Ring (middle), and the Lower Ring (bottom), as shown in the image on the left. Black squares indicate the observed data with associated uncertainties. Emission lines in the H band are modeled using \texttt{lmfit}, while spectral features in the Y band are fit using \texttt{pPXF}. The dashed lines mark the observed wavelengths of H$\alpha$ and Ca\,\textsc{ii}~H at the redshift of the Host galaxy, serving as a visual reference for identifying relative blueshifts or redshifts in other regions.}
    \label{fig:Slit_Spec}
\end{figure*}

We use \texttt{prospector} \citep{2021Johnson} to model the SEDs of individual regions, adopting the FSPS stellar population library \citep[][]{2009Conroy,2010Conroy}, MIST isochrones \citep[][]{2016Choi}, and the C3K spectral library \citep[][]{2020Cargile}, with a Chabrier IMF \citep{2005Chabrier}. The model incorporates 22 free parameters that describe stars, gas, and dust. For a detailed discussion of the \texttt{prospector} models used here, we refer to \cite{2024Park} and the references therein. 
The nonparametric SFH is modeled using 14 logarithmically spaced bins, with a Student-T prior applied to $\log\,\mathrm{SFR}$ \citep{2019Leja}. Two SFH priors are used: the continuity prior ($\sigma = 0.3$, $\nu = 2.0$) to reduce sharp fluctuations, and the bursty prior ($\sigma = 1.0$, $\nu = 1.0$) for more flexibility. The two priors yield consistent SFH and SFR estimates, so only results from the continuity prior are presented for clarity.

In Fig.~\ref{fig:region_sfh} the SFH of each region is shown in a color-coded box which is connected to the corresponding region by an arrow. The estimated stellar mass $\mathrm{M_\star}$ (listed in Table~\ref{tab:Mass_RA_Dec}) and optical dust attenuation $\mathrm{A_v}$ for each region are displayed within the respective boxes. The associated uncertainties represent the $\mathrm{1\sigma}$ confidence intervals around the maximum-a-posteriori estimates. 

The Host galaxy and the Inner Ring exhibit clear signatures of a recent starburst and significant presence of dust. The Host galaxy has an estimated total stellar mass of $\mathrm{10^{11.4}\,M_\odot}$ (see Table~\ref{tab:Mass_RA_Dec}) and $\mathrm{10^{10.7\pm0.2}\,M_\odot}$ in the remained central region (navy region in Fig.~\ref{fig:system}) with an $\mathrm{A_v}$ of $\mathrm{0.7\pm1.3}$. The Inner Ring shows a higher dust content with $\mathrm{A_v}=2.1\pm1.0$, and together with the Lower Ring contains about half of the total stellar mass of the Host galaxy. Both the Host and the Inner Ring appear to have experienced a starburst event at roughly the same time, leading to similar star formation rates of $\mathrm{\sim 300 ~M_\odot yr^{-1}}$. The apparent similarity in SFH between these two regions may be partly due to contamination of the Inner Ring photometry by the bright Host galaxy. 

\begin{table*}[ht]
    \centering
    \resizebox{0.8\textwidth}{!}{
    \begin{tabular}{c|ccc}
\hline
\diagbox[width=3.5cm,height=3.5em]{\hspace{-40pt}Region/Object}{\hfill Vel [$\mathrm{km\,s^{-1}}$]} & H Band (\textsc{lmfit}) & Y Band (\textsc{PpXF}) & Prospector (phot+spec)\\

\hline
Upper Ring & $91.8\pm6.6$ & -- & -- \\
\hline
Bullet & -- & $-92.0\pm30.2$ & $-55.4\pm12.9$ \\
\hline
Host & $0.0\pm5.4$ & $0.0\pm51.5$ & $0.0\pm8.9$ \\
\hline
Lower Ring & $-66.8\pm11.1$ & -- & $-57.1\pm11.3$ \\
% \hline

    \end{tabular}
    }
    \caption{Relative velocities of four distinct components in the system, reported with respect to the Host galaxy. Velocities derived from the H  and Y band spectra are based on redshift differences in emission and absorption lines, respectively. The \texttt{Prospector}-based velocities correspond to the best-fit model redshift parameters obtained by jointly fitting the photometric and spectroscopic data, as described in Sec.~\ref{spectral_analysis}.}
    \label{tab:velocities}
\end{table*}

The SFH of the Bullet, on the other hand, shows evidence of rapid quenching over a timescale of $\mathrm{<0.2-0.3 Gyr}$. The inferred stellar mass from the SED is $\mathrm{10^{11.1\pm0.2}\,M_\odot}$, which is approximately 0.7 times the combined mass of the Host and its associated rings (see Tab.~\ref{tab:Mass_RA_Dec}), and about 2.5 times more massive than the central region of the Host galaxy (i.e., the dark blue region). The $\mathrm{A_v}$ on the other hand is extremely low for the Bullet compared to its companion. Interestingly, the SFH of the Bullet galaxy shows no clear evidence of a starburst, even by employing bursty SFH (see Sec.~\ref{sec:photometry}). This absence may be an important clue to the physical processes involved in quenching the Bullet, as we discuss in Sec.~\ref{sec:discussion}.

The Upper Ring exhibits a more complex SFH in comparison to the Lower Ring, which may be partly due to flux contamination from the Bullet and/or the tail, because both the colors and the ionized gas emission lines (analyzed in Sec.~\ref{spectral_analysis}) suggest that the Upper and Lower Ring have similar SFR. The tidal tail itself is characterized by an old stellar population and signs of recent quenching, with a significantly low SFR ($\sim 10^{-1.5} \ M_\odot \, yr^{-1}$), suggesting a cold-gas poor environment consistent with expectations \citep[e.g.,][]{2018Renaud}. We note that modest changes to the segmentation boundaries do not introduce significant differences in the derived stellar masses or star formation histories, with results remaining consistent within the quoted uncertainties.

\section{Spectroscopic Analysis}\label{spectral_analysis}

Figure~\ref{fig:Slit_Spec} presents the Keck/MOSFIRE spectra. The left panel displays the RGB composite image of the system, overlaid with the actual footprint of the spectroscopic slit.
The slit covers the tidal tail, the Upper Ring, the Bullet, the Host galaxy, part of the Inner Ring, and the Lower Ring, providing valuable information about the entire system. We define three color-coded regions: the upper region (red box) captures the contribution from both the Bullet galaxy and the upper part of the ring. The middle region (blue box) encompasses the Host galaxy along with a portion of the Inner Ring, while the lower region (green box) primarily traces the emission from the Lower Ring.
We separate the two-dimensional spectra into the three distinct regions, and extract a Y-band and an H-band spectrum for each region.
Due to the seeing, the regions actually observed are slightly wider than the slit. The three regions are then identified in the imaging data, and the corresponding SEDs are extracted.  

The spectra for each region are shown in the right panels of Figure~\ref{fig:Slit_Spec}, in color-coded boxes. The H-band spectra show prominent $\mathrm{H\alpha}$, $\mathrm{[N\,II]}\,\lambda6548\,\text{\AA}$, and $\mathrm{[N\,II]}\,\lambda6584\,\text{\AA}$ emission lines from ionized gas. The Y-band spectra include mostly stellar absorption features such as Balmer and Ca lines.

\begin{figure*}
    \centering
    \includegraphics[width=1.\linewidth]{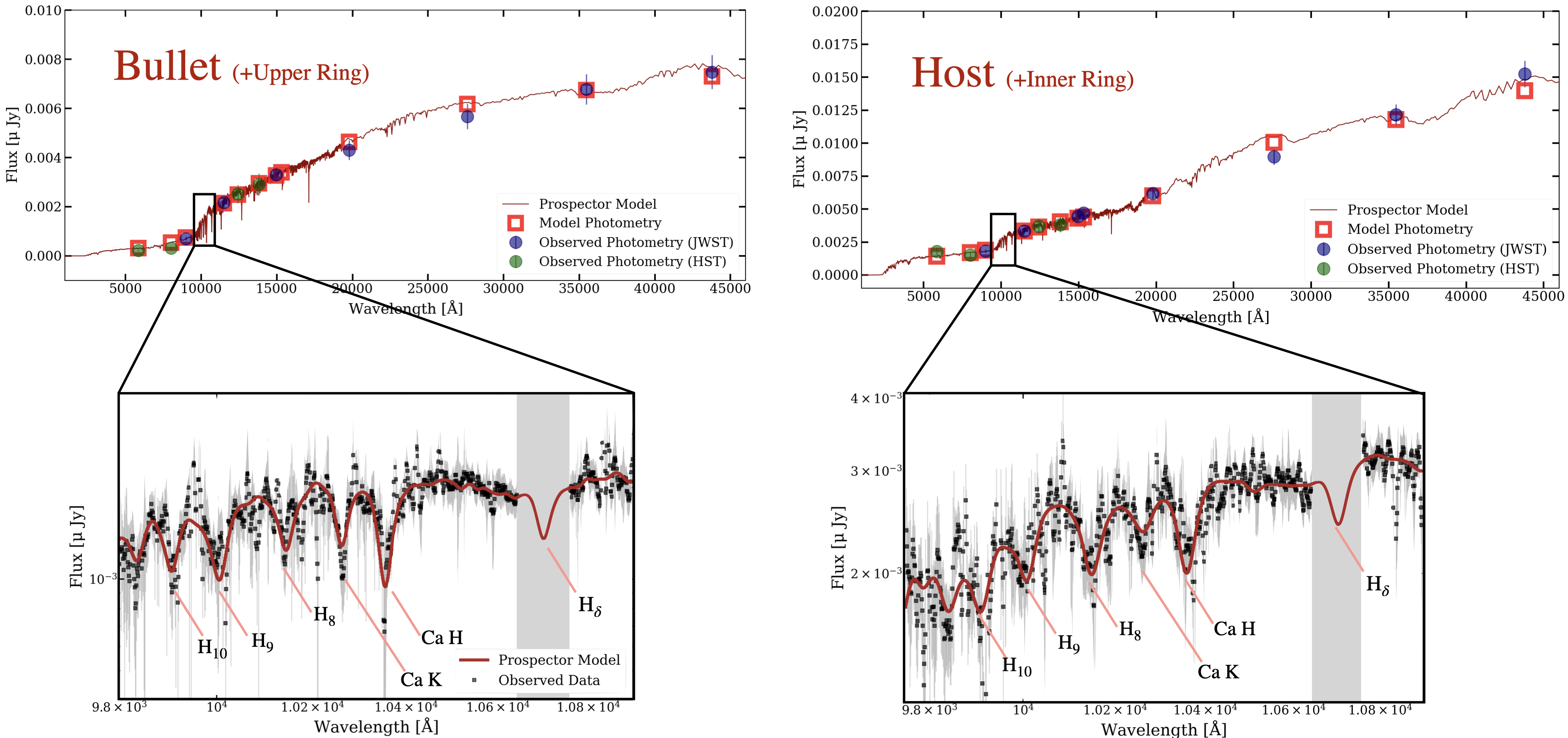}
    \caption{The best-fit SED models of the Host and Bullet galaxies, derived using \texttt{\textsc{Prospector}}. The fits incorporate photometric data from HST and JWST/NIRCam (top panels), along with Y-band spectra from MOSFIRE (bottom panels), as described in Sec.~\ref{sec:SFH}. The wavelength range covered by the spectra is magnified for both objects and overlaid with the corresponding best-fit spectral models. Grey shaded regions in the zoomed-in panels indicate masked intervals containing $\mathrm{H{\delta}}$ emission lines. } 
    \label{fig:Fit_Prospect_Appendix}
\end{figure*}

\subsection{Gas Kinematics}\label{subsec:kinematics}

In order to analyze the kinematics of the ionized gas, we fit the emission lines in the H-band spectra using the \texttt{LMFIT} package \citep{2016Newville}, performing the fits independently for each region. The continuum is modeled using a second-order polynomial fit to the spectral wavelength window. The emission lines are modeled as Gaussians, all sharing a common velocity and dispersion ($\sigma$), which are fixed to those of the narrow H$\alpha$ line. To account for the presence of the AGN and the apparent line broadening in the Host spectrum, we include an additional broad component to the H$\alpha$ line. This broad component is allowed to have a central velocity offset of up to $\pm200$ km\,s$^{-1}$ relative to the narrow component and is constrained to have $\sigma < 5000$ km\,s$^{-1}$. To quantify the uncertainties associated with the line fluxes, we adopt a Monte Carlo simulation technique. This involves perturbing the flux values at each wavelength point 200 times by sampling from a Gaussian distribution, where the mean corresponds to the measured flux and the standard deviation is set by the flux error.

The spectrum of the central region containing the Host galaxy exhibits a prominently broadened H$\alpha$ emission line, with a broad component velocity dispersion of $\sigma \sim 2100$ km\,s$^{-1}$ and a narrow component of $\sigma = 213 \pm 19$ km\,s$^{-1}$. This, combined with the observed narrow line ratio of [N\,\textsc{ii}]/H$\alpha \simeq 1$, is characteristic of AGN emission \citep[see, e.g.,][]{2001Kewley,2003Kauffmann} and  is consistent with the classification of the Host galaxy as an X-ray detected AGN (see Sec.~\ref{sec:intro}). The broad $\mathrm{H\alpha}$ component is negligible in both the upper and lower regions. 

The fit results indicate that the narrow component of the emission lines has a different velocity in each of the three regions.
This is evident in Figure~\ref{fig:Slit_Spec}, where the vertical dashed line marks the position of the narrow $\mathrm{H\alpha}$ line in the Host galaxy: compared to this reference, the upper region exhibits a redshift of about 90~km/s, whereas the lower region shows a blueshift of about $-70$~km/s. The corresponding projected velocity offsets are summarized in Table~\ref{tab:velocities}.

The top region includes both the Bullet and the Upper Ring, but we attribute the observed emission lines to the Upper Ring, because the velocity dispersion and flux ratio measurements are consistent with those obtained for the Lower Ring in the bottom region. Moreover, the redshift of the emission lines in the top region does not match the redshift of the stellar absorption lines in the same region, as discussed below. We thus conclude that in this region we are seeing the continuum (in both the H and Y band) mostly coming from the Bullet, and the emission lines mostly coming from the Upper Ring.

Having measured the kinematics of the Upper and Lower Ring, we can now attempt to determine the rotational and radial (expansion) velocities of the collisional ring. This can be done by analyzing the line-of-sight (LOS) velocity as it varies with deprojected position angle (PA) at a constant radius along the ring (see Fig.~\ref{fig:appx_geo}). We estimate the ring diameter to be approximately 19.2 kpc (see Appendix~\ref{sec:geometry}), with an inclination angle of $\mathrm{17_{-3}^{+5}}$ degrees, derived from the observed ring width of $\sim$1.6 kpc. The inclination allows for deprojecting the observed LOS velocities. Serendipitously, the lower part of the MOSFIRE slit includes the intersection of the Lower Ring with the system’s major axis (see Appendix \ref{sec:geometry}), where the LOS velocity predominantly reflects rotational motion (see Fig.~\ref{fig:appx_schema}). This alignment enables us to derive a deprojected rotational velocity of $\mathrm{228_{-52}^{+57}}$ km/s, fully compatible with the rotational velocity expected for the measured stellar mass \citep[e.g.,][]{2020Cannarozzo}. Using the average angular offset of the Upper Ring from the system’s measured PA, estimated to be approximately $55^\circ$, we disentangle two perpendicular velocity components—radial (expansion) and rotational—in the Upper Ring (see Fig.~\ref{fig:appx_schema}). Given this geometry, the rotational contribution to the observed H$\alpha$ velocity (see Table~\ref{tab:velocities}) is estimated to be $\mathrm{42_{-10}^{+11}}$ km\,s$^{-1}$. Subtracting this from the measured LOS velocity yields the projected expansion velocity. Accounting for the inclination of the system, we derive a deprojected radial (expansion) velocity of $\mathrm{127_{-29}^{+72}}$ km\,s$^{-1}$.

\subsection{Stellar Kinematics}\label{subsec:stellarkinematics}

We measure the stellar kinematics in each region using \texttt{pPXF} \citep{Cappellari2017} to model the absorption lines in the Y-band spectra. The stellar continuum is modeled with 32 simple stellar population templates from the EMILES library \citep{Vazdekis2016}. These templates are based on the \cite{Chabrier2003} initial mass function, BaSTI isochrones \citep{Pietrinferni2004}, and include eight stellar ages spanning from 0.15 to 14 Gyr, logarithmically spaced in steps of 0.22 dex. Four metallicity values were considered: [Z/H] = –1.5, –0.35, 0.06, and 0.4. During the fitting process, we applied multiplicative polynomials of sixth order to account for variations in the spectral shape. Note that the H$\delta$ emission line was fit with a velocity allowed to vary within $\pm200$\,km\,s$^{-1}$ relative to the stellar component, to accommodate potential kinematic offsets.

In the central region, the stellar kinematics indicates a velocity offset of $67 \pm 79 \, \mathrm{km\,s^{-1}}$ relative to the redshift measured from the emission lines, confirming that both the stellar and gas components trace the same system, the Host galaxy. In the lower region the Y-band continuum is faint, as it originates solely from the Lower Ring. The absence of identifiable absorption features in this spectrum prevents us from deriving a reliable redshift based on absorption lines.
The top region, instead, reveals a markedly different velocity in the Y band compared to the H band (see Table~\ref{tab:velocities}), indicating that the continuum and the emission lines likely originate from distinct components. While the emission lines are redshifted compared to the Host, the absorption lines are blueshifted. Given that the Bullet galaxy is quiescent and red, we attribute the Y-band continuum to this galaxy, while the strong ionized gas emission seen in the H band is likely due to the Upper Ring which is star-forming---consistent with expectations from extensive studies of CRGs (see Sec.~\ref{sec:intro}).

In conclusion, analysis of the kinematics shows that the Bullet galaxy is moving toward the observer, while the Upper Ring is gas-rich and moving away from the observer. While the Bullet and the Upper Ring appear connected in the imaging, this must be clearly a projection effect, with the Bullet likely being in the foreground of the Upper Ring.

\subsection{\texttt{Prospector} Fit and SFHs}\label{sec:SFH}

\begin{figure}
    \centering
    \includegraphics[width=1.\linewidth]{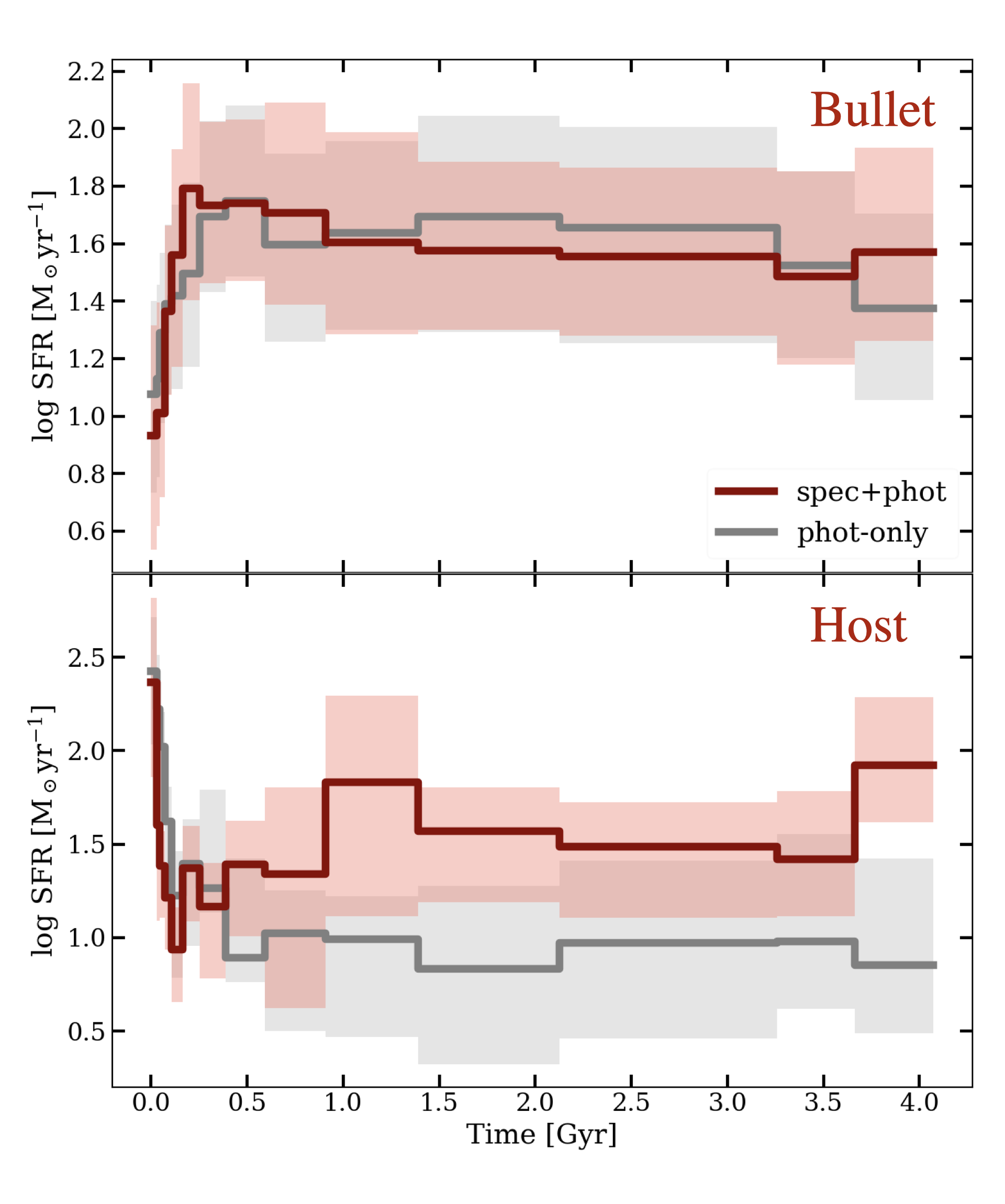}
    \caption{ Star formation histories of the Bullet (top panel) and the Host (bottom panel) galaxies derived using JWST/NIRCam and HST photometry, along with Y band MOSFIRE spectroscopy, modeled with \texttt{\textsc{Prospector}}. In each panel, the SFH inferred from photometry alone is shown in gray, while the combined photometry plus Y band spectroscopy result is shown in dark red. Shaded regions indicate the 1$\sigma$ confidence intervals around the maximum a posteriori estimates in each age bin.}
    \label{fig:SFH}
\end{figure}

\begin{figure*}[ht]
    \centering
    \includegraphics[width=1.\linewidth]{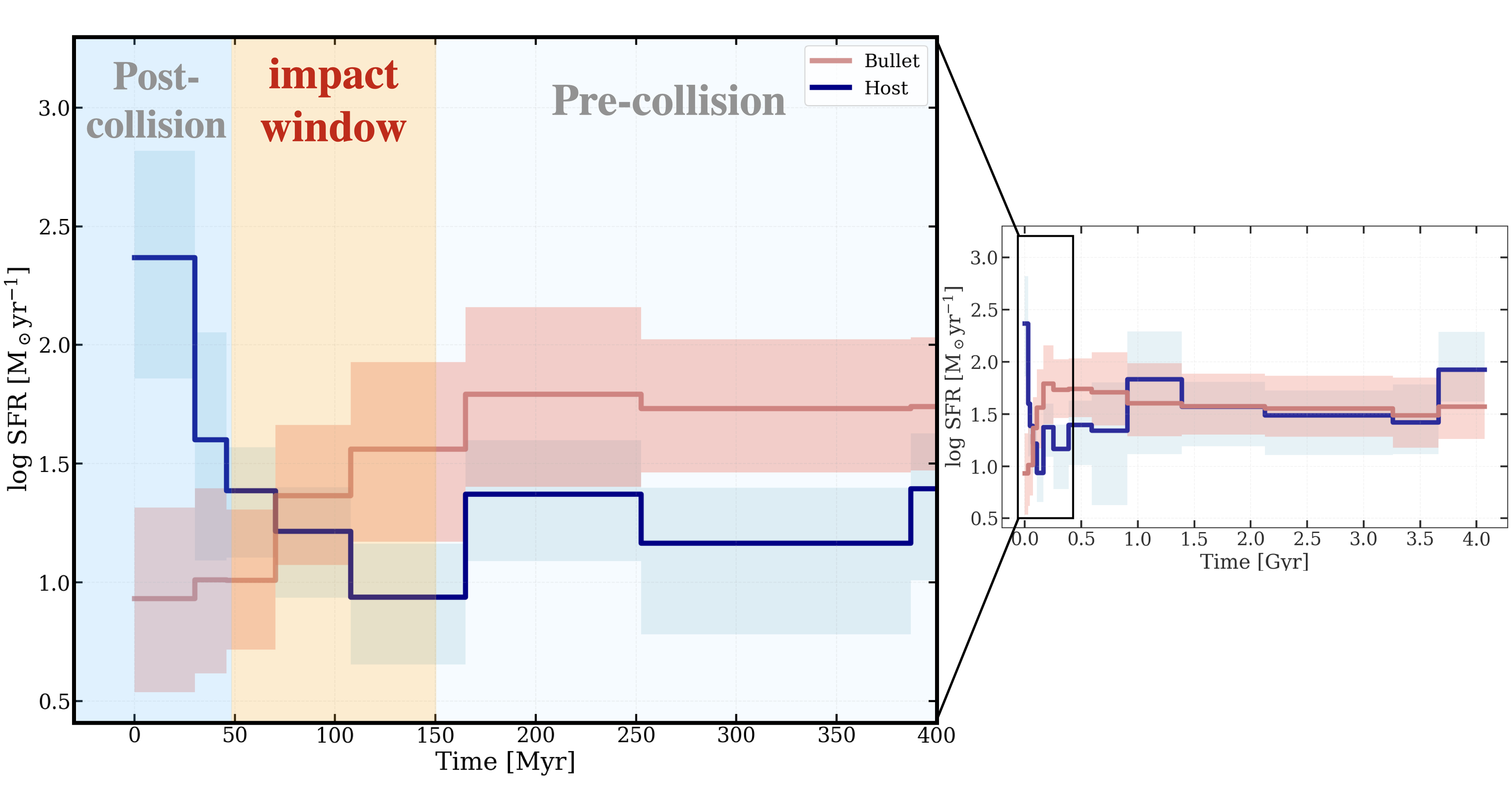}
    \caption{Star formation histories of the Bullet and the Host galaxy derived using JWST/NIRCam and HST photometry, along with Y band MOSFIRE spectroscopy, modeled with \texttt{\textsc{Prospector}}. The left panel is a zoomed-in SFH in the last 6 age bins ($< 400$ Myr). Shaded orange region indicates the impact window where we expect the first impact took place.}
    \label{fig:SFH_zh}
\end{figure*}

Similarly to what done with the photometry and explained in Sec.~\ref{sec:photometry}, we use \texttt{prospector} to model the SED of different regions, but this time including both photometric data and Y-band spectra. The model includes 29 free parameters that describe the stars, gas, and dust, as explained earlier, with additional parameters necessary to fit the spectroscopy, accounting for example for imperfect flux calibration and spectral uncertainties, as explained in detail by \cite{2024Park}.
It is important to note that while dust emission is considered in our models, AGN and nebular emission are not included in our current model. For this reason, we mask the H$\delta$ emission lines in the spectra and do not include the H-band spectra in the fit.

The best-fit \texttt{Prospector} model and the observed data are presented in Fig.~\ref{fig:Fit_Prospect_Appendix} for the Bullet and Host galaxies. We fit MOSFIRE Y-band spectra in combination with JWST and HST photometric data, extracted from the same spatial regions. Notably, the Host spectrum appears to exhibit a small excess absorption at the Ca K line (CaII $\lambda3969$\AA), which is not fully accounted for by the stellar population template. This excess is also visible in Fig.~\ref{fig:Slit_Spec}, where the Y band spectrum is modeled using \texttt{pPXF}. The Ca K line has attracted growing attention in recent literature \citep[][]{2024Belli,2025Liboni} because it is a resonant transition and thus can be used to detect cold gas in the galaxy reservoir and/or outflows, as an alternative diagnostic to the Na~D line. The gas traced is mostly neutral, given that the ionization potential of Ca II (11.9 eV) is lower than that of hydrogen.

In line with the objectives of this study, Fig.~\ref{fig:SFH} presents the SFHs of the Bullet and Host derived using two methods: (1) combined photometry and spectroscopy, and (2) photometry alone. For both methods, the photometric data were extracted from identical spatial regions, corresponding to the Keck boxes shown in Fig.~\ref{fig:Slit_Spec} (distinct from those defined in Sec.~\ref{sec:photometry}). The results demonstrate a high degree of consistency between the two approaches, with both yielding comparable present-day SFRs (first age bin) and similar burst/quenching patterns. Noticeable discrepancy appears in the older age bins of the Host galaxy’s SFH. However, this variations do not affect the overall interpretation or conclusions of this work. Note that the associated uncertainties of the derived SFR in each age bin represent the $1\sigma$ confidence intervals around the maximum a posteriori estimates.

\section{Discussion}\label{sec:discussion}

\subsection{Quenching During the Impact Window}

The kinematics of the collisional ring enables a reconstruction of the time elapsed since the impact between the two galaxies (see Sec.~\ref{subsec:kinematics}). We assume that the ring started expanding at the time of impact and that the expansion rate was constant (see also \citealt{2018Renaud} and references therein). Accounting for observational uncertainties, the estimated collision time ranges from 47 to 96 Myr ago. This measurement is consistent with the ring detectability range \citep[$<500$ Myr, ][]{2018Renaud,2023AMartinez-Delgado} and observed rings lifetime in previous studies at lower redshifts \citep[$\sim$ 50 to 300 Myr, ][]{2011Conn,2024Ditrani,2012Smith}. The 47 Myr lower bound corresponds to the maximum inclination and minimum rotational velocity. Recent observational \citep[e.g.,][]{2012Scudder,2013Patton} and simulation studies \citep[e.g.,][]{2016Sparre,2018Renaud} suggest that changes in the SFH of interacting galaxies can begin well before first contact or final coalescence (at separations $<30$ kpc), sometimes as early as $\sim 50$ Myr prior to impact. Incorporating this pre-collision influence, we estimate a window of approximately 50-150 Myr over which the interaction can impact the SFH. 

The timeline of the impact is compared to the SFH of the Host and the Bullet in Fig.~\ref{fig:SFH_zh}.
As illustrated by the shaded regions in the figure, the impact window timeline ($\sim$50–150 Myr) aligns closely with the onset of major changes in star formation activity—specifically, a pronounced starburst in the Host galaxy and rapid quenching in the Bullet. This temporal correspondence highlights the pivotal role of the merger in driving these evolutionary transitions. Quantitatively, the Host exhibits a star formation rate more than 25 times higher than at the onset of the impact window, while the Bullet’s star formation rate declines to nearly one-tenth of its pre-impact value, demonstrating contrasting responses by the two galaxies to the same interaction event. As discussed by \citep{2018Renaud}, variations in the SFH of collisional systems typically peak 50–150 Myr after the interaction, consistent with the rapid evolution observed at t = 0 in our system (see Fig.~\ref{fig:SFH}).    

The alignment between the timeline of quenching and that of the galaxy interaction strongly suggests the presence of a physical connection between these two processes. However, in principle it is possible that this is merely a coincidence, and the Bullet was instead quenched by an unrelated mechanism while it was also interacting with the Host.
The likelihood of a massive galaxy undergoing quenching independently of an external impact can be roughly estimated. \citet{2023Park} provide number density estimates for recently quenched galaxies at redshift $z \approx 1.6$ with stellar masses comparable to our system. They report that the number density of galaxies that quenched rapidly within the past 300 Myr at this redshift is approximately $2 \times 10^{-6}~\mathrm{Mpc^{-3}~Gyr^{-1}}$. According to \citet{2023Weaver}, based on COSMOS2020 data, the total number density of galaxies with comparable stellar mass at $z \approx 1.6$ is about $7 \times 10^{-4}~\mathrm{Mpc^{-3}~Gyr^{-1}}$, with quiescent galaxies comprising roughly $2 \times 10^{-4}~\mathrm{Mpc^{-3}~Gyr^{-1}}$ of that population. This means that if we select a massive quiescent galaxy at $z\sim1.6$ that is in a major merger, the probability that this system is undergoing quenching within this short timescale purely by chance—i.e., independently of the interaction with the companion—is of the order of $1\%$. This low probability strengthens the case for merger-induced quenching in our system.

\subsection{Alternative Process Associated with Galaxy Mergers}

Quenching in merging systems is typically believed to occur due to the alteration of gas inflow, typically by rapidly consuming the gas reservoir of the galaxy during significant bursts of star formation \citep[e.g.,][]{2022Ellison,2023Li}.  However, we rule out the possibility that quenching in the Bullet is caused by gas exhaustion based on the absence of any detectable starburst signature in the SFH (see Figs.~\ref{fig:region_sfh} and \ref{fig:SFH}). This conclusion holds even when adopting a bursty SFH model in \texttt{Prospector} (see Sec.~\ref{sec:photometry}).
Another possibility is that the interaction triggers an episode of strong AGN feedback in the Bullet, which may rapidly remove the ISM gas content and shut down the star formation. In this classic scenario \citep[e.g.,][]{2005DiMatteo,2008Hopkins}, the black hole accretion is also typically accompanied by a substantial starburst, which we do not detect.
In contrast, the Host galaxy exhibits both a significant starburst and AGN activity, indicating substantial gas inflow during the collision event. This suggests a gas-rich environment, where gas is funneled into the central regions as a result of the impact.

Several studies have proposed that galaxy interactions can induce significant turbulence within the ISM, which in turn may suppress star formation efficiency by disrupting the gravitational collapse of gas clouds \citep[e.g.,][]{2018Ellison,2018vandeVoort,2025Suess}. This phenomenon is sometimes associated with so-called violent disc instability, a process that can be triggered during major mergers \citep[e.g.,][]{2014Dekel&Burkert,2015Zolotov}. In the context of our system, such a scenario remains plausible, particularly if the timescale for gas inflow is shorter than the timescale required for star formation. This condition corresponds to a wetness parameter\footnote{This parameter is defined as the ratio of the timescale required for star formation over gas inflow timescale.} exceeding unity, favoring inflow-dominated evolution over immediate star formation. Notably, in our case, the turbulence and instability within the ISM are more likely driven by the dynamical effects of the merger itself, rather than by starburst-induced feedback, which is minimal or absent in the quiescent companion galaxy. Future investigations employing spatially resolved studies of turbulent and shocked regions will be crucial for developing a deeper understanding of this scenario.

\subsection{The Dragon Effect: AGN-Driven Negative Feedback on Companion Galaxies}

While AGN and starburst activities are not detected in the Bullet, they are clearly apparent in the Host. The presence of ongoing AGN feedback in the Host galaxy is confirmed by the X-ray detection along with the presence of a broad H$\alpha$ emission line. We thus explore the idea that the AGN in a galaxy may exert negative feedback on the companion galaxy, a possibility that has been discussed in recent studies  \citep{2020Davies,2024Perez-Gonzalez,2025Balashev,2025Suzuki}. We refer to this mechanism as the \textit{Dragon Effect}: a process by which star formation is effectively quenched in the companion galaxy due to AGN-driven outflow from its neighbor. This disruption can delay or suppress future star formation for extended periods, representing a novel form of external feedback in galaxy evolution and a new angle in environmental quenching at cosmic noon, where the peak of mergers, quenching, and AGN activity occurs. This scenario is particularly relevant given that the number density of AGN is known to peak in merging systems \citep[e.g.,][]{2005Springel,2025Ellison,2025Euclid}, suggesting that such interactions—and their feedback effects—may be more common and influential than previously thought.

In a recent study, \citet{2025Balashev} report the discovery of a major merging system at redshift $z \approx 2.7$, where they show that radiation from a quasar hosted by one of the galaxies directly impacts the ISM of its companion. Their analysis reveals that in regions exposed to the quasar's radiation, low-density molecular gas is significantly disrupted, leaving behind only highly compact ($>10^5 \, \mathrm{cm^{-3}}$ $\mathrm{H_2}$ number density), dense clouds that are insufficient to sustain star formation. The effect of the quasar’s intense UV radiation on molecular gas is strikingly similar to the disruptive influence of massive newborn stars on their natal clouds, fundamentally altering the internal gas structure and likely suppressing star-forming activity. Additionally, a statistical analysis by \citet{2025Suzuki} of quasar environments at \( z \sim 2.2 \), irrespective of alignment with quasar outflow trajectories, reveals a notable deficit of Ly\(\alpha\) Emitters (LAEs) within quasar proximity regions compared to continuum-selected galaxies. The suppression is most pronounced for LAEs with high Ly\(\alpha\) equivalent widths, and both LAE and continuum galaxy number densities decline with decreasing distance to the quasar. These findings support the interpretation that quasar radiation can photoevaporate gas in low-mass halos—which typically host LAEs—and may even impact more massive halos, thereby suppressing star formation across a range of galaxy masses.

Such a mechanism represents a form of external, environment-induced quenching, in which feedback from a neighboring galaxy halts or delays star formation in a companion system. We note that the Host features some evidence of neutral gas, detected as an excess absorption in Ca~K, which may be associated with a dense neutral outflow. Future investigations with spatially resolved data tracing molecular clouds and neutral gas distributions will be crucial to test and better understand this scenario.

\section{Summary and Conclusion}\label{conclusion}

In this work, we present the discovery and analysis of a high-redshift collisional ring galaxy system at $\mathrm{z = 1.61} $ in the UDS field, identified through visual inspection of publicly available \textit{JWST}/NIRCam imaging. The system comprises a massive AGN-hosting ring galaxy with \( M_\star \sim 10^{10.7}\, M_\odot \) in the central region, and a similarly massive companion—the Bullet—with \( M_\star \sim 10^{11.1}\, M_\odot \). The ring exhibits a well-defined axisymmetric structure, bright spokes, and an extended tidal tail—morphological features consistent with a major merger origin.

By combining deep photometry with Keck/MOSFIRE H  and Y band spectroscopy, we extract spatially resolved kinematics and SFHs for multiple regions within the system. From the kinematics of the deprojected ring ($\sim 20$~kpc in diameter), we infer a rotational velocity of \( \mathrm{228^{+57}_{-52}} \) km\,s\(^{-1} \) and an expansion velocity of \( \mathrm{127^{+72}_{-29}} \) km\,s\(^{-1} \), leading to an estimated collision time between 47 and 96 Myr prior to observation. Accounting for potential pre-impact star formation changes, we infer a total SFH altering interaction window of approximately 50–150 Myr. This timeframe closely matches the observed starburst in the Host and the quenching in the Bullet, supporting a direct causal link between the merger and the divergent evolutionary states of the two galaxies.
Given the low probability ($<1\%$) that rapid quenching would occur in the Bullet galaxy at the observed time without any external influence, our findings strongly favor a model in which the merger is physically responsible for the quenching of star formation. 

We evaluate several potential quenching mechanisms for the Bullet galaxy, including merger-induced turbulence, violent disc instability, and AGN-driven negative feedback. Based on the absence of a recent starburst in the Bullet, we rule out gas depletion due to starburst activity as the primary quenching driver. Instead, the spatial geometry and dynamical signatures point to a merger-driven interaction, where gravitational and kinematic disturbances induce turbulence in the Bullet’s ISM, potentially suppressing star formation by preventing the collapse of gas into bound structures.

In addition, we propose the \textit{Dragon Effect}---a negative feedback mechanism in which AGN-driven outflows from a rapidly accreting galaxy suppress star formation in a nearby companion. This scenario is supported by recent observations that show signs of disrupted low-density molecular gas and altered gas properties near powerful AGNs. The Dragon Effect thus represents a compelling new pathway for environmental quenching of massive galaxies during cosmic noon, with broader implications for understanding the interplay between galaxy interactions and feedback processes.

This system thus provides the first direct observational connection between a major merger timeline and the onset of bursting and quenching phases in massive galaxies. More broadly, it offers a rare case study in how major mergers and AGN feedback can jointly regulate star formation during cosmic noon---an epoch marked by the peak of galaxy assembly, star formation, and black hole growth. Further observations of the distribution of molecular gas in such systems, along with spatially resolved investigations of their ISM properties, can shed light on the physics behind massive galaxy quenching and help determine the dominant physical mechanisms at play. Additionally, future multi-phase hydrodynamic simulations will be able to fully assess the Dragon Effect by quantifying its dependence on AGN power, impact parameter, and the gas structure of the companion.

\section{Data Availability}
{ \bf Some of the data presented in this article were obtained from the Mikulski Archive for Space Telescopes (MAST) at the Space Telescope Science Institute. The specific observations analyzed can be accessed via \dataset[DOI: 10.17909/yxwb-6d35]{https://doi.org/10.17909/yxwb-6d35}}.

% \section{Acknowledgments}
\begin{acknowledgments}

AK, SB, LB, and MS are supported by the the ERC Starting Grant “Red Cardinal”, GA 101076080. This work is based on observations made with the NASA/ESA/CSA James Webb Space Telescope. The data were obtained from the Mikulski Archive for Space Telescopes at the Space Telescope Science Institute, which is operated by the Association of Universities for Research in Astronomy, Inc., under NASA contract NAS 5-03127 for JWST. These observations are associated with program GO 1817. This work also makes use of observations taken by the 3D-HST Treasury Program (GO 12328) with the NASA/ESA HST, which is operated by the Association of Universities for Research in Astronomy, Inc., under NASA contract NAS5-26555.

\end{acknowledgments}

\bibliography{ref.bib}{} 
\bibliographystyle{aasjournal}

\appendix

\section{Fate of The System}\label{Appx:fate of the system}

\begin{figure*}[ht]
    \centering
    \includegraphics[width=.8\linewidth]{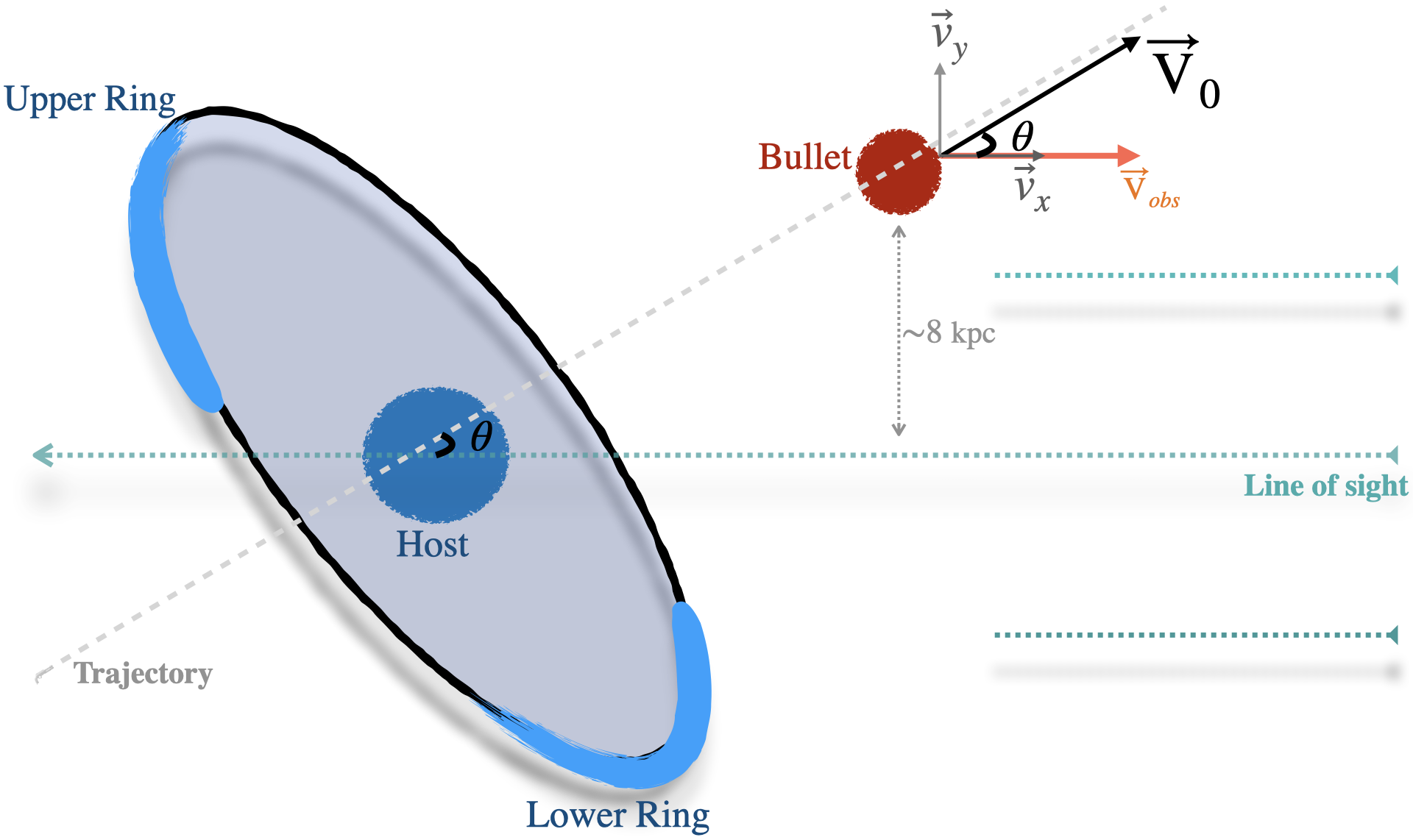}
    \caption{Schematic representation of the Bullet and the ring system in a plane perpendicular to the LOS. The relative velocity of the Bullet with respect to the Host, denoted as $V_0$, has both LOS and perpendicular components in $v_x$ and $v_y$ velocity space. In this configuration, the LOS component corresponds to the observed velocity difference between the Bullet and the Host (see Table~\ref{tab:velocities}). The angle $\theta$ represents the inclination between the Bullet’s trajectory and the LOS direction.}
    \label{fig:appx_Fate}   
\end{figure*}

To evaluate the future evolution of the system using observational data, we need to estimate both the relative distance and velocity between the Bullet and the Host galaxies. The presence of a CRG provides a crucial constraint: the trajectory and impact position are naturally centered near the host galaxy’s nucleus, as typical for these systems (see Fig.~\ref{fig:appx_Fate}). Considering the nature of CRG, we can assume the angular momentum, L, is negligible (see Sec.~\ref{sec:intro}). In our case, the projected separation between the Bullet and the center of the fitted ellipse (Fig.~\ref{fig:appx_geo}, left panel) is approximately $\sim 8$ kpc. The LOS velocity difference between the Bullet and Host is measured to be $\sim 90$ km/s (see Table~\ref{tab:velocities}).

Considering the geometry of the system as illustrated in Fig.~7, and indicating with subscript $0$ the quantities at the time of observation, we can write
\begin{equation}
    \xzero = \Dzero \cos\theta,    
\end{equation}
\begin{equation}
    \yzero = \Dzero \sin\theta,    
\end{equation}
\begin{equation}
    \vxzero = \vzero \cos\theta,    
\end{equation}
\begin{equation}
    \vyzero = \vzero \sin\theta,    
\end{equation}
where $x$ is along the line of sight, $y$ is orthogonal to the line of sight, and $0 < \theta < \pi/2$. Here, $\Dzero$ and $\vzero$ are the intrinsic separation and relative speed, respectively. We have measurements of $\yzero \approx 8\ \mathrm{kpc}$ and $\vxzero \approx 90\ \mathrm{km\,s^{-1}}$.

A key constraint on the   dynamics comes from the inferred impact time, estimated at $t_0 \approx 47–96$~Myr ago from the expansion of the collisional ring (see Sec.~\ref{sec:discussion}). Over this interval, the Bullet must have traveled roughly $8\ \mathrm{kpc}$ perpendicular to the LOS, i.e., along the $y$ direction. We then make the simple assumption that the relative velocity of the galaxies has remained constant during the $\sim$100 Myr since the impact. This yields an estimate of $\vyzero = \yzero / t_0 \approx 160$~km/s. In reality, the velocity of the Bullet decreases as it moves away from the Host, so that our estimate is an upper limit for the true value of $\vyzero$. We thus calculate an upper limit on the total relative velocity of $\mathrm{V_{B} \approx 120~km\,s^{-1}}$, corresponding to $\theta \sim 34^\circ$. This, in turn, implies a lower limit on $\Dzero$ of $\sim 12.5$ kpc.

%Using the projected separation of $\sim 8\ \mathrm{kpc}$, we can place constraints on the inclination angle $\theta$ between the Bullet’s trajectory and the LOS—a critical parameter for reproducing the observed kinematics—rather than directly measuring it.

The orbital energy per unit mass in the point-mass approximation is
\begin{equation}\label{E_eq_zero}
    E = \frac{1}{2} \vzero^2 - \frac{G M_{\mathrm{tot}}}{\Dzero},
\end{equation}
where $M_{\mathrm{tot}}$ is the sum of the masses of the two galaxies (including both baryonic and dark matter components).  
Since the separation $\Dzero$ ($\mathrm{>12.5\ kpc}$) is not particularly large—especially near the upper limit of $\theta$—compared to the system’s size ($3$–$4\ \mathrm{kpc}$, the typical half-mass radius at this redshift and stellar mass; \citealt{2010van_Dokkum}), we adopt a conservative assumption that only the stellar masses contribute to $M_{\mathrm{tot}}$ (i.e., $M_{\mathrm{tot}} = M_{\star,\mathrm{tot}}$), neglecting the contribution of the dark matter halo. Although this approach underestimates the absolute value of the gravitational potential energy, it ensures that the point-mass approximation remains sufficiently accurate as long as the stellar mass distributions of the two galaxies have minimal overlap at their current separation. 

Using $\yzero = \Dzero \sin\theta$ and $\vxzero = \vzero \cos\theta$, Eq.~\ref{E_eq_zero} can be rewritten as
\begin{equation}\label{E_eq_theta}
    E = \frac{1}{2} \frac{\vxzero^2}{\cos^2\theta} - \frac{G\,M_{\mathrm{tot}}\sin\theta}{\yzero}.
\end{equation}

With $M_{\mathrm{tot}} = M_{\star,\mathrm{tot}}$, we find $E < 0$ for $\theta > 2^\circ$. Given that systems with $L=0$ and $E < 0$ are expected to merge rapidly \citep{1987Binney}, this implies that for $\theta > 2^\circ$ the system will undergo a rapid merger. For $\theta = 2^\circ$ the current relative velocity is $\vzero \approx \vxzero$, the current separation is $\Dzero = 220\,$kpc and the mean relative speed (computed as current separation divided by time elapsed  since the collision) is $\mathrm{V_{mean}}\approx2100\, \mathrm{km\,s^{-1}}$. A smaller value of $\vzero$ and larger values of $\Dzero$ and $\mathrm{V_{mean}}$ would be found for $\theta < 2^\circ$. The inferred $\mathrm{V_{mean}}$ implies a relative speed at the time of collision far exceeding $2100\, \mathrm{km\,s^{-1}}$—a value highly implausible even under extreme conditions and well beyond observed limits \citep[e.g.,][]{1995Marzke,1999Struble,2004Markevitch,2004Jing}. Therefore, given the energy of the system, we conclude that the system is very likely to coalesce rapidly.

\section{Geometry}\label{sec:geometry}

\begin{figure*}[ht]
    \centering
    \includegraphics[width=.94\linewidth]{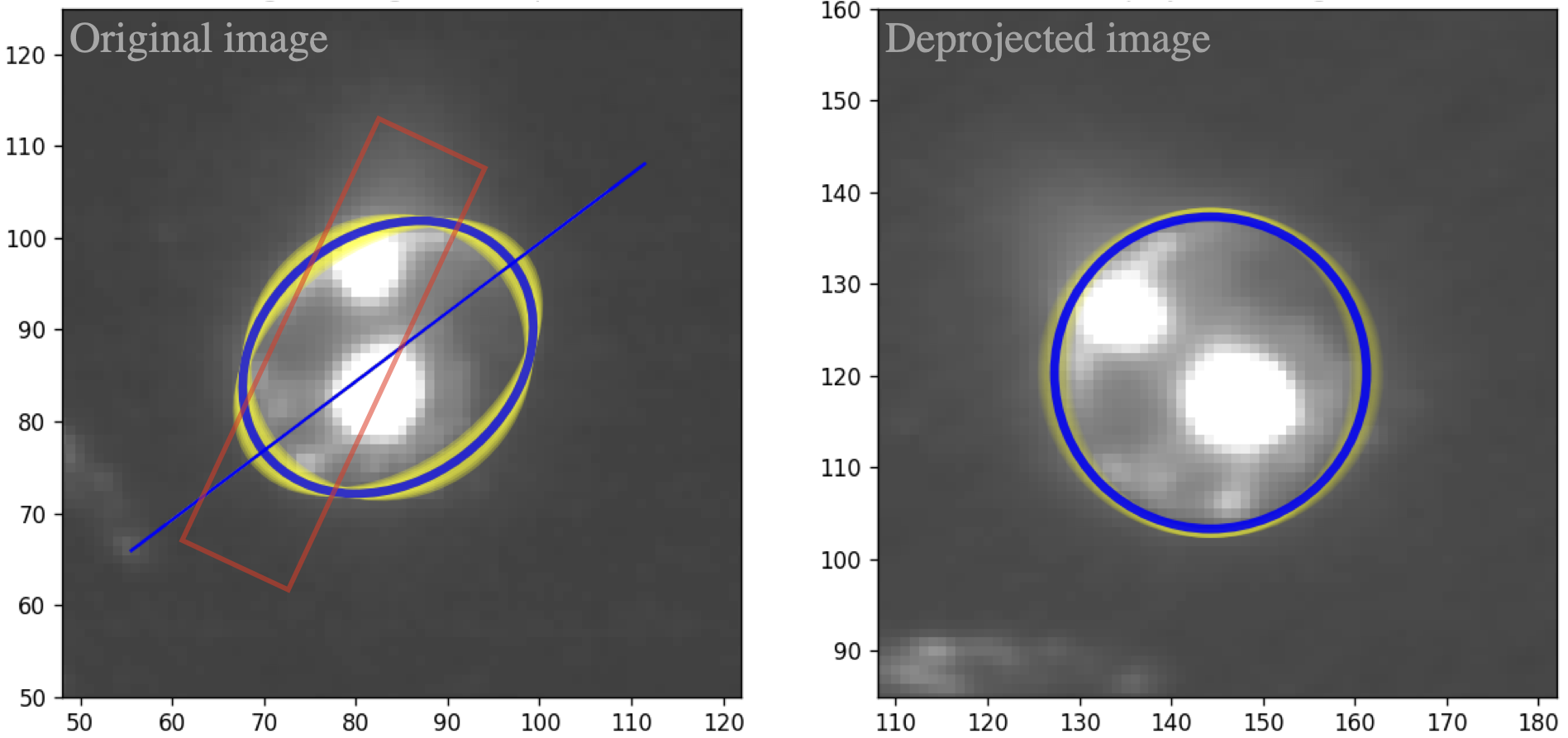}
    \caption{The original and deprojected images are presented here, with blue lines indicating the best-fit ellipses and yellow lines representing the uncertainty in the inclination. The original image is a composite constructed from all NIRCam broad-band filters used in this study (see Section~\ref{sec:photometry}). The MOSFIRE slit is shown as a red rectangle overlaid on the image.}
    \label{fig:appx_geo}
\end{figure*}

We utilize JWST/NIRCam imaging in the F090W, F115W, F150W, and F200W filters to reconstruct the projected geometry of the ring system. To begin, we define the major axis of the ring as the largest distance between the two farthest points along the ring's edge, an axis that remains unaffected by inclination. Based on this, we identify the perpendicular (minor) axis by placing a cross-sectional box across the ring and measuring the peak surface brightness, after masking out foreground/background sources and the bright spokes. This procedure yields an initial estimate of the ring’s apparent axial ratio, from which the inclination and position angle are derived. These geometric parameters are visualized with the blue ellipse in Fig.~\ref{fig:appx_geo} (left panel).

Assuming the intrinsic shape of the ring is circular, we deproject the ellipse to reconstruct the true geometry, as shown in the right panel of Fig.~\ref{fig:appx_geo}. To estimate the ring’s width, we again mask the spokes and sources and place four rectangular apertures oriented radially, covering the bright arcs in both the Upper and Lower Ring regions. The ring width is then defined as the FWHM of the surface brightness profile along each aperture. From the measured ring thickness and ellipticity, we refine the inclination estimate by sampling across multiple positions within the ring. The variation in inclination measured from different apertures is used to define the uncertainty in the final inclination and, by extension, the deprojection of the system.

\begin{figure*}[ht]
    \centering
    \includegraphics[width=.9\linewidth]{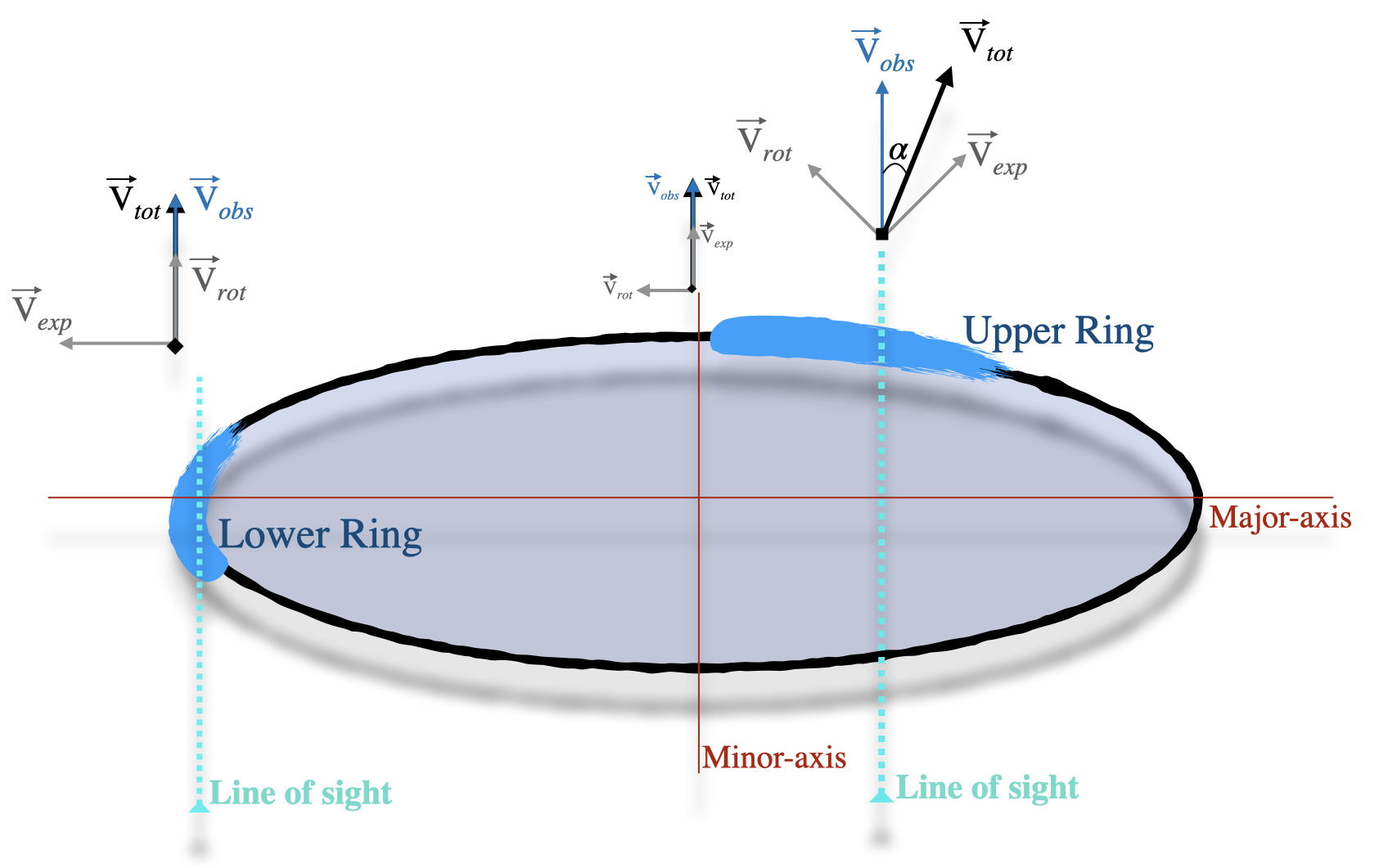}
    \caption{Schematic representation of the ring system, with the position angle neglected for simplicity. The total velocity is defined as the vector sum of the rotational and expansion components: $\vec{V}_{\mathrm{tot}} = \vec{V}_{\mathrm{rot}} + \vec{V}_{\mathrm{exp}}$. In this configuration, the LOS component of the velocity corresponds purely to expansion along the minor axis and purely to rotation along the major axis. The angle $\alpha$ denotes the separation between the LOS direction and the orientation of $\vec{V}_{\mathrm{tot}}$.}
    \label{fig:appx_schema}
\end{figure*}

In Fig.~\ref{fig:appx_schema} we show a diagram of the ring system to illustrate the relation between the observed LOS velocity of the gas in each point and its rotational and expansion velocity components, which is discussed in Sec.~\ref{subsec:kinematics}.

\end{document}